\documentstyle[epsfig,12pt,a4wide]{article}
\def\fc#1#2{\frac{#1}{#2}}
\def\g{\gamma}
\def\d{\delta}
\def\csl#1{\slash\!\!\!\! #1}

\def\11{11-{\it d} supergravity}

\def\a{{\mathcal{A}}}
\def\an{ansatz }
\def\h#1{\hat #1}
\def\t#1{\tilde #1}
\def\b#1{\bar #1}
\def\w{\wedge}

\def\bbbone{{\mathchoice {\rm 1\mskip-4mu l} {\rm 1\mskip-4mu l}
{\rm 1\mskip-4.5mu l} {\rm 1\mskip-5mu l}}}

\def\eps{\epsilon}
\def\na#1{\nabla_{#1}}

\def\bbbc{{\mathchoice {\setbox0=\hbox{$\displaystyle\rm C$}\hbox{\hbox
to0pt{\kern0.4\wd0\vrule height0.9\ht0\hss}\box0}}
{\setbox0=\hbox{$\textstyle\rm C$}\hbox{\hbox
to0pt{\kern0.4\wd0\vrule height0.9\ht0\hss}\box0}}
{\setbox0=\hbox{$\scriptstyle\rm C$}\hbox{\hbox
to0pt{\kern0.4\wd0\vrule height0.9\ht0\hss}\box0}}
{\setbox0=\hbox{$\scriptscriptstyle\rm C$}\hbox{\hbox
to0pt{\kern0.4\wd0\vrule height0.9\ht0\hss}\box0}}}}

\newcommand\bbbR{{\mathrm {I\!R}}}

\begin{document}

\setcounter{page}{0}

\thispagestyle{empty}
\begin{flushright}
DAMTP-2000-137 \\
\end{flushright}
{\LARGE 
\begin{center}
{\scshape M-Theory~on~Seven~Manifolds~\\with~$G$-fluxes}
\end{center}}
\vspace{.1in}
\begin{center}
{\large
Tibra Ali\footnote{T.Ali@damtp.cam.ac.uk}\\
\vspace{.1in}}
\emph{D.A.M.T.P.\\
Centre for Mathematical Sciences\\ 
University of Cambridge\\
Wilberforce Road, Cambridge CB3 0WA, UK}
\end{center}

\vspace{.2in}
\begin{abstract}
We compactify M-theory on seven-manifolds with a warp-factor and
$G$-fluxes on the internal space. Because of non-zero $G$-fluxes, we
are forced to adopt a Majorana supersymmetry spinor ansatz which does
{\it not} have the usual direct product structure of two lower
dimensional Majorana spinors. For the spinor
ansatz that we choose, we find that supersymmetry puts strong
constraints on the internal space namely that it must be conformal to a Ricci-flat
seven-manifold of the form $X^7= X^6 \times X^1$. The holonomy of
$X^6$ must be larger than $\bbbone$ if the warp-factor is to be
non-trivial. The warp-factor depends only on the $X^1$
direction and is singular. We argue that to avoid this singularity one
has to embed this solution in a Ho\v rava-Witten setup and thus has natural links to much studied brane-world scenarios.
\end{abstract}
\newpage

\setcounter{footnote}{0}


\section{Introduction}
\setcounter{equation}{0} 
The purpose of this paper is to examine
the role of warp-factors and $G$-fluxes in supersymmetric
compactifications of M-theory on Ricci-flat 7-manifolds so that the
resulting theory has 4-{\it d} Minkowski slices ($\bbbR^{1,3}$). M-theory,
which is yet to be fully understood, admits as it's low energy limit
\11 \cite{Cremmer}. In four-dimensions spinors can not be chiral and
Majorana simultaneously. This subtlety becomes relevant when
compactifying supergravity theories down to four dimensions with
warp-factors that preserve some fraction of supersymmetry. In some
recent works \cite{Acharya,Hernandez,Becker2} on warped compactifications of M-theory on
seven-manifolds this seems to have been overlooked.
\par
For our spinor \an we adopt the one proposed in \cite{deWit} and find as a consequence that our internal
manifold has to be conformal to a Ricci-flat 7-manifold of the form
$X^7=X^6\times X^1$. Therefore our internal space  is not a {\it bona
fide} 7-manifold like a $G_2$-holonomy space, which has received a lot
of attention recently, but our calculation points to certain pitfalls
one has to circumvent if one is to compactify on 7-manifolds.
\par
A few years ago it was shown by Becker and Becker \cite{Becker} that
when we turn on certain components of the background $G$-fluxes in M-theory
compactified on a 8-manifold, supersymmetry requires that internal
space to be a Calabi-Yau 4-fold. The external space of this solution is
$\bbbR^{1,2}$. This work suggests that it might be possible to do the
something similar with $\bbbR^{1,3}$ and an internal Ricci-flat
manifold.
\subsection*{\it Some historical facts about compactification on 7-manifolds}
An older review of the subject is \cite{Duff1}. Perhaps, the most famous
solution is the one by Freund and Rubin \cite{Freund}. By choosing an \an such that $G$ is
proportional to the volume-form of the four-dimensional Lorentzian
subspace, one obtains a solution of the form $AdS^4\times X^7$, where
$X^7$ is a 7-{\it d} positive curvature Einstein space. For maximal
supersymmetry $X^7$ is usually taken to be $S^7$. This
has been dubbed ``spontaneous compactification''. It is not possible
to flatten the $AdS^4$ without decompactifying the internal $S^7$. This is not the
only problem besetting spontaneous compactification. Another problem is
the absence of chiral multiplets. This shortcoming was rectified in
the heterotic version of M-theory by Ho\v rava and Witten
\cite{Horava}. It has been realised recently that compactifying on
$G_2$-holonomy spaces with some special singularities can
also be used to generate chiral multiplets \cite{Acharya2}. But for the question we
are interested in the most interesting solution is that of \cite{deWit}. The solution presented in
\cite{deWit} is a warped product of $AdS^4$ and $S^7$. Their \an for
the $G$-fluxes is like that of Englert \cite{Englert}. The Englert \an
consists of, in addition to a Freund-Rubin piece, turning on those
components of $G$ which live exclusively on the internal
manifold. These are the only components that can be turned on if one
is to have a maximally symmetric space-time as the external space. The
Englert solution is not in general supersymmetric. In \cite{deWit} the
$G$-fluxes are related to the torsion on $S^7$ via the Maxwell
equation and the resulting
theory has $G_2$-invariance and thus $N=1$ supersymmetry. The claim is
that this solution corresponds to the $G_2$-invariant point of
the superpotential of the 4-{\it d} $N=8$ gauged supergravity theory
of de Wit and Nicolai \cite{deWit2}. The important point of their
solution is that their 11-{\it d} Killing spinor, $\eta$, is not a
direct product of 4- and 7-dimensional Majorana spinors. This subtlety does not arise in the simplest $S^7$
compactifications and compactifications on manifolds with dimension
other than seven\footnote{\, See for example \cite{Becker}.}. While
the usual spinor \an $\eta=\eps\otimes\theta$, with $\eps$ and
$\theta$ are 4-{\it d} anticommuting and 7-{\it d} commuting Majorana
Killing spinors, simplify life considerably, it restricts the space of
supersymmetric solutions to those without warp-factors.
\par
It has been argued in \cite{Gibbons2,Maldacena} that the solutions put
forward in \cite{Becker,Dasgupta} do not exist when one considers the
compactification on {\it compact} manifolds. However, it was also
noted in \cite{Maldacena} that this type of solutions can exist when
the M5-brane anomaly term is added to the \11 action
\cite{Duff2}. 
\subsection*{\it Outline}
We now present an outline of the sections of
our paper. The next section of this paper is to fix our
conventions. In section 3 we present our main calculations related to
supersymmetry and obtain constraints on the compactification space and
the warp-factor. In section 4 we give a geometric interpretation of
the supersymmetry constraints and show that the internal space is
conformal to $ X^6\times X^1$. The reducibility of the internal space
is seen as a consequence of the Killing spinor \an that we have
chosen. What emerges from this discussion is
that the warp-factor can have non-trivial dependence only along $X^1$. We then solve Einstein's equation to find the shape of the
warp-factor which unfortunately turns out to be singular. We then close with some comments about
what could be the application of such a solution. In the appendix we discuss the four-seven split of
11-{\it d} Majorana spinor and give some Clifford algebra identities used
in section 3.

\section{Metric Ansatz and Conventions}
\setcounter{equation}{0}
The bosonic part of the action for \11 is \cite{Cremmer}
\begin{eqnarray}
{\mathcal{S}}_{11}=\fc{1}{2}\int d^{11}x
\sqrt{-\b{g}}\left.\b{R}\right.-\int\left[\fc{1}{4}\b{G}\w \b*\b{G}+\fc{1}{12}\b{C}\w \b{G}\w
\b{G}\right] \label{eq:Met1},
\end{eqnarray} 
where the first term is the Einstein-Hilbert term and $\b C$ is the
3-form potential and $\b G=d\b C$ is the corresponding 4-form field
strength. The only fermionic matter field in this theory is the gravitino, $\psi_{M}$, which is
a spinor-vector, whose
variation under supersymmetry is given by
\begin{equation}
\delta_{\eta}\psi_{M}={\bar{\nabla}}_{M}\eta-{\bar{Z}}_{M}\eta \label{eq:Met2},
\end{equation}
where $\eta$ is the Majorana supersymmetry parameter and
\begin{equation}
\bar{Z}_{M}=\fc{1}{288}\left[{{\bar{\Gamma}}_{M}\!\!}^{PQRS}-8{\delta_{M}}^{P}\bar{\Gamma}^{QRS}\right]\b{G}_{PQRS}
\label{eq:Met3}
\end{equation}
\begin{equation}
\bar{\Gamma}^{{M_1}{M_2}...{M_n}}=\bar{\Gamma}^{[M_1}\bar{\Gamma}^{M_2}...\bar{\Gamma}^{M_n]}.
\label{eq:Met4}
\end{equation}
If a bosonic
configuration is to preserve supersymmetry then $\eta$ must satisfy
the Killing spinor equation,
\begin{equation}
{\bar{\nabla}}_{M}\eta-{\bar{Z}}_{M}\eta=0 \label{eq:Met5}.
\end{equation}
\par
We adopt the usual warp-factor metric \an
appropriate for $X_4\times X_7$,
\begin{eqnarray}
\bar{g}_{MN}(x,y)=D^{-1}(y)\left[ \begin{array}{cc}
g_{\mu\nu}(x)  & 0 \\
0 & g_{mn}(y) \end{array} \right]=D^{-1}(y)g_{MN}(x,y) \label{eq:Met6}
\end{eqnarray}
where we follow the convention that $x^{\mu}$ ($y^{m}$) denote
the coordinates of the 4-dimensional (7-dimensional) space
whose geometry is described by the metric $g_{\mu\nu}$
($g_{mn}$).
\par
We want to write out the Killing spinor equation in terms of the
unbarred metrics and so we observe
\begin{eqnarray}
{\bar{\nabla}}_{M}=\na{M}-\fc{1}{4}D^{-1}{\Gamma_{M}}^{N}\left(\na{N}D\right)
\\
\begin{array}{cc}
\bar{\Gamma}_{M}=D^{-\fc{1}{2}}\Gamma_{M} &
\bar{\Gamma}^{M}=D^{\fc{1}{2}}\Gamma^{M}\end{array}. \label{eq:Met7}
\end{eqnarray}
We shall assume that $\bar{G}_{MNPQ}$ has zero conformal
weight. Discussions on Clifford algebra identities and Majorana
spinors relevant for the calculations are given in the appendix.

\section{The $\bbbR^{1,3}$ Solution with $G$-fluxes}
\setcounter{equation}{0}
The most general choice of $G$ consistent with compactifications down to
maximally symmetric 4-{\it d} space-time is given by the Englert \an \cite{Englert}:
\begin{equation}\begin{array}{rcl}
G_{\mu\nu\rho\sigma}&=&6m\eps_{\mu\nu\rho\sigma} \\
G_{mnpq} &\not=&0.
\end{array}
\end{equation}
For our solution we shall ultimately set the Freund-Rubin parameter
$m$ to zero. With the appropriate choice for the Dirac matrices (see
appendix \ref{ap:a} for details) the
four-dimensional part of (\ref{eq:Met5}) now becomes,
expressed in terms of the unbarred metric,
\begin{eqnarray}
\na{\mu}\eta-\fc{\partial_{m}(\log{D})}{4}\left(\gamma_{\mu}\Sigma^{5}\otimes\gamma^{m}\right)\eta
-\fc{D^{\fc{3}{2}}}{288}\left(\gamma_{\mu}\otimes\csl{G}\right)\eta
-im D^{\fc{3}{2}}\left(\gamma_{\mu}\Sigma^{5}\otimes\bbbone\right)\eta=0
\label{eq:Killing1}
\end{eqnarray}
where  $\csl{G}\equiv G_{mnpq}\gamma^{mnpq}$
and
$\Sigma^{5}\equiv\fc{i}{4!}\eps_{\mu\nu\rho\sigma}\g^{\mu}\g^{\nu}\g^{\rho}\g^{\sigma}$
is the 4-{\it d} chirality operator. Now if we assume $\na{\mu}\eta=0$ for Minkowski space then we are left
with the condition
\begin{eqnarray}
\fc{\partial_{m}(\log{D})}{4}\left(\Sigma^{5}\otimes\gamma^{m}\right)\eta
+\fc{D^{\fc{3}{2}}}{288}\left(\bbbone\otimes\csl{G}\right)\eta
+ im
D^{\fc{3}{2}}\left(\Sigma^{5}\otimes\bbbone\right)\eta=0 \label{eq:cond}
\end{eqnarray}
The integrability for (\ref{eq:Killing1}) for $\bbbR^{1,3}$ yields the
following condition
\begin{eqnarray}
\left[\fc{1}{16}\partial_{m}(\log{D})\partial^{m}(\log{D})-D^{3}m^2\right]\left(\bbbone\otimes\bbbone\right)\eta
\nonumber \\
-\fc{D^{3}}{(288)^{2}}\left(\bbbone\otimes\csl{G}^2\right)\eta+\fc{D^{\fc{3}{2}}\partial_{m}(\log{D})}{144}\left(\Sigma^{5}\otimes
G^{m}\right)\eta   \label{eq:Integ} \\
+\fc{imD^{\fc{3}{2}}\partial_{m}\left(\log{D}\right)}{2}\left(\bbbone\otimes\gamma^{m}\right)\eta=0
\nonumber
\end{eqnarray}
where $G^{m}\equiv\g_{pqr}G^{mpqr}$. In the absence of a warp-factor this reduces to
\begin{equation}
\fc{1}{(288)^2}\left(\bbbone\otimes\csl{G}^2\right)\eta=-m^2\left(\bbbone\otimes\bbbone\right)\eta.
\end{equation}
This equation implies that $\eta$ is an eigenspinor of $\csl{G}^2$
with negative eigenvalue. Since  $\csl{G}$ is a hermitian operator the
eigenvalues of $\csl{G}^2$ are positive semi-definite. So the above
equation implies
\begin{equation}\begin{array}{rcl}
m&=&0\\
\csl{G}\eta&=&0. \label{eq:Candelas}
\end{array}
\end{equation}
This implies, via the internal-space Killing spinor equation
integrability condition, that
all components of $G$ has to vanish together with the fact that the
internal space has to be Ricci-flat \cite{Candelas}.
\par
When there is a non-constant warp-factor present and $m=0$ (\ref{eq:Integ}) becomes
\begin{eqnarray}
\fc{\partial_{m}(\log{D})\partial^{m}(\log{D})}{16}\left(\bbbone\otimes\bbbone\right)\eta-\fc{D^{3}}{(288)^{2}}\left(\bbbone\otimes\csl{G}^2\right)\eta
+\fc{D^{\fc{3}{2}}\partial_{m}(\log{D})}{144}\left(\Sigma^{5}\otimes
G^{m}\right)\eta=0
\end{eqnarray}
This equation is satisfied if $\eta$ is a simultaneous eigenspinor
of $\fc{D^{3}\;\csl{G}^{2}}{(288)^{2}}$ and  $\fc{D^{\fc{3}{2}}\partial_{m}(\log{D})}{144}\left(\Sigma^{5}\otimes
G^{m}\right)$ with eigenvalues
${\partial_{m}(\log{D})\partial^{m}(\log{D})}\over{16}$ and
zero, respectively. These requirements are satisfied if
\begin{equation}
-\fc{\partial_{n}(\log{D})}{24}\left(\bbbone\otimes{\gamma_{m}}^{n}\right)\eta
=\fc{D^{\fc{3}{2}}}{288}\left(\Sigma^{5}\otimes G_{m}\right)\eta
\label{eq:cond2}.
\end{equation}
It is easy to see that this condition implies the one suggested by
(\ref{eq:cond}) with $m=0$,
\begin{equation}
-\fc{\partial_{n}(\log{D})}{4}\left(\Sigma^{5}\otimes\gamma^{n}\right)\eta
=\fc{D^{\fc{3}{2}}}{288}\left(\bbbone\otimes\csl{G}\right)\eta \label{eq:cond1}.
\end{equation}
The Clifford algebra identities given in the appendix are useful in these manipulations.
\subsection*{\it Structure of the Killing Spinor}
What does (\ref{eq:cond1}) tells us about the structure of the Killing
spinor? It was noted in \cite{deWit} that turning on $G$-fluxes forces
us to consider Killing spinors which can not be of the form
$\eta=\eps\otimes\theta$, where $\eps$ and $\theta$ are Majorana and
pseudo-Majorana Killing spinors on $\bbbR^{1,3}$ and the internal space $X^7,$ respectively. We can
see that very easily from the
foregoing conditions. Let us suppose, for the sake of argument, that the
Majorana Killing spinor has the form
\begin{equation}
\eta=\eps\otimes\theta \label{eq:supp1}.
\end{equation}
We assume $\theta$ to be pseudo-Majorana and commuting. Since $\eta$
is Majorana and anticommuting, $\eps$ must be Majorana and
anticommuting. Then suppose that (\ref{eq:cond1}) is satisfied by
\begin{eqnarray}
-\fc{\partial_{n}(\log{D})}{4}\g^{n}\theta=\fc{D^{\fc{3}{2}}}{288}\csl{G}\theta=\theta'
\label{eq:supp2}
\end{eqnarray}
for some $\theta'$. We now have two cases. If $\theta'\not=0$ then we arrive at the contradiction
\begin{eqnarray}
\Sigma^5\eps=\eps.
\end{eqnarray}
If on the other hand $\theta'=0$ then by squaring the left hand side
of (\ref{eq:supp2}) we obtain
\begin{equation}
\partial_{m}(\log{D})\partial^{m}(\log{D})=0
\end{equation}
which means that the warp-factor is constant and we are left with a
special case of (\ref{eq:Candelas}) and hence $G=0$. This complication doesn't
occur, for example, in compactification on 8-manifolds \cite{Becker}
where one can define the internal spinors to be chiral and Majorana
simultaneously. So the only way to
introduce non-trivial $G$-flux and warp-factor is to introduce a
non-minimal structure of $\eta$. Following \cite{deWit} we make the
\an
\begin{equation}
\eta=\left[A+B_{m}\left(\Sigma^{5}\otimes\gamma^{m}\right)\right](\eps\otimes\theta)\equiv
\a (\eps\otimes\theta)
\label{eq:ansatz}
\end{equation}
where $A$ and $B_{m}$ are assumed to be real functions of
$y^{m}$. Reality is necessary for $\eta$ to a Majorana spinor.
\par
We insert our \an (\ref{eq:ansatz}) into (\ref{eq:cond2}) and since
$\eps$ and $\Sigma^5\eps$ are linearly independent, we obtain
\begin{equation}
-\fc{D^{\fc{3}{2}}}{288}\left.G^{m}B_{p}\g^{p}\theta=\fc{\partial_{n}(\log{D})A}{24}\right.
\g^{mn}\theta
\label{eq:cond2a}
\end{equation}
and
\begin{equation}
-\fc{D^{\fc{3}{2}}A}{288}\left. G^{m}\theta=\fc{\partial_{n}(\log{D})}{24}\right.\left[B^{p}{\g^{mn}}_{p}\theta-B^{m}\g^{n}+B^{n}\g^{m}\right]\theta
\label{eq:cond2b}.
\end{equation}
Contracting with $\g_m$ we obtain
\begin{equation}
-\fc{A}{4}\partial_{m}\left(\log{D}\right)\g^{m}\theta=\fc{D^{\fc{3}{2}}}{288}\left[B_{m}\g^{m}\csl{G}-8B_{m}G^{m}\right]\theta
\label{eq:cond1a}
\end{equation}
and
\begin{equation}
\fc{\partial_{n}\left(\log{D}\right)}{4A}\left[B_{m}\g^{mn}-B^{n}\right]\theta=\fc{D^{\fc{3}{2}}}{288}\left.
\csl{G}\theta \right. \label{eq:cond1b}.
\end{equation}
These condition are of course the conditions that we get by putting
our \an into (\ref{eq:cond1}).
\par
Let us now see the effect of (\ref{eq:cond2}) on the internal space
Killing spinor equation. This equation is
\begin{equation}
\na{m}\eta-\fc{1}{4}\partial_{n}\left(\log{D}\right)\left(\bbbone\otimes{\gamma_{m}}^{n}\right)\eta-\fc{D^{\fc{3}{2}}}{288}\left[\Sigma^{5}\otimes{\gamma_{m}}^{pqrs}G_{pqrs}-8\Sigma^{5}\otimes G_{m}\right]\eta=0
.
\end{equation}
By using 
\begin{eqnarray*}
{\gamma_{m}}^{pqrs}G_{pqrs}=\gamma_{m}\;\csl{G}-4G_{m}
\end{eqnarray*}
we get
\begin{equation}
\na{m}\eta-\fc{\partial_{n}(\log{D})}{4}\left(\bbbone\otimes{\gamma_{m}}^{n}\right)\eta-\fc{D^{\fc{3}{2}}}{288}\left[\Sigma^{5}\otimes\gamma_{m}
\csl{G}-12\Sigma^{5}\otimes
G_{m}\right]\eta=0.
\end{equation}
By using (\ref{eq:cond2}) we obtain
\begin{equation}
\na{m}\eta+\fc{1}{4}\partial_{m}\left(\log{D}\right)\left(\bbbone\otimes\bbbone\right)\eta-\fc{1}{2}\partial_{n}\left(\log{D}\right)\left(\bbbone\otimes{\gamma_{m}}^{n}\right)\eta=0.
\end{equation}
Now defining
\begin{eqnarray}
\tilde{\eta}&=&D^{\fc{1}{4}}\eta \\
\hat{g}_{mn}&=&D^{-2}g_{mn} \label{eq:newmetric}
\end{eqnarray}
we obtain
\begin{equation}
\hat{\nabla}_{m}\tilde{\eta}=0. \label{eq:G2}
\end{equation}
This equation ensures that the internal space is conformal to a
Ricci-flat manifold with holonomy contained within $G_2$. To obtain a
nontrivial solution we demand that
$\t{\theta}=D^{\fc{1}{4}}\theta$ be covariantly constant with
respect to $\h{g}_{mn}$. So we get
\begin{equation}
(\h\nabla_{m}\a)(\eps\otimes\t\theta)=0.
\label{eq:condansatz}
\end{equation}
The solution to (\ref{eq:condansatz}) consistent
with the conditions (\ref{eq:cond1a}-\ref{eq:cond1b}) is
\begin{equation}
A={\mathrm Constant.}
\end{equation}
and
\begin{equation}
B_{m}=\pm\fc{\partial_{m}(\log{D})}{\sqrt{\partial_{n}(\log{D})\partial^{n}(\log{D})}}.
\label{eq:B}
\end{equation}
where ${\bf{\hat{\nabla} \hat{B}}}=0$.

\section{Geometric Interpretation}
If ${\mathcal X}$ is a holonomy-${\mathcal H}$ Riemannian manifold and
there exists on ${\mathcal X}$ a tensor field ${\bf T}$ such that 
\begin{equation}
{\bf \nabla T}=0 \label{eq:condT}
\end{equation}
then it is a fundamental result of the theory of holonomy groups
\cite{Lichnerowicz} that ${\bf T}$
must be invariant under the ${\mathcal H}$. I.e.
\begin{equation}
\Lambda {\bf T}={\bf T}\;\forall\Lambda\in {\mathcal H}.
\end{equation}
\par
Let us apply this result to our case. Let us denote by $X^7$ the
Ricci-flat manifold described by the metric defined in
(\ref{eq:newmetric}), which dropping the hat, we denote from now on as
$g_{mn}$. In this section we work with this metric and we drop all
hats from all formulae. Equation (\ref{eq:G2}) implies that
it's holonomy group, ${\mathcal H}\subset G_2$. Since ${\bf
B}$ is a vector field then we can write it locally as
\begin{equation}
{\bf B}=\fc{\partial}{\partial z}.
\end{equation}
If the tangent space splits up under the action of the holonomy group
locally then this implies via a fundamental theorem in
differential geometry \cite{Guggenheimer} that $X^7=X^6 \times
X^1$. It also means that ${\bf B}$ can only be a function of the
coordinate of $X^1$ which we denote by $z$. Therefore of all the
possibilities we must exclude the ones with ${\mathcal H}=G_2$.
\par
One
way of understanding the exclusion of $G_2$ solutions is as
follows. $G_2$-holonomy spaces admit only one commuting pseudo-Majorana
spinor, $\tilde\theta$ \cite{Joyce,Gibbons,Salamon}. However since $\bar{\tilde\theta} \g^{m} \tilde\theta=0$, there is
no covariantly constant vector associated to this spinor. And we
have seen that for our ansatz supersymmetry demands such a covariantly constant
vector. It seems to us that to obtain a $G_2$-holonomy solution one
should adopt a spinor ansatz of the form
\begin{equation}
\eta=\left[A+B_{pqr}\left(\Sigma^{5}\otimes\gamma^{pqr}\right)\right](\eps\otimes\theta).
\end{equation}
We wish to investigate
this ansatz further in a future communication.
\par
Since ${\bf B}$ is unit norm (\ref{eq:condT}) does not put any more constraints on the warp-factor. So
to determine the shape of the warp-factor we have to solve the bosonic
equations which we do in the next section. Our task has been simplified since the warp-factor can
depend only on $z$.
\par
Let us now enumerate the different possibilities for ${\mathcal H}$. To
this end we define a 3-form
\begin{equation}
\phi_{mnp}=i\;\bar{\t{\theta}}{\g}_{mnp}\t{\theta} \label{eq:3form}
\end{equation}
and so its Hodge-dual with respect to $g_{mn}$ is given by
\begin{equation}
*\phi_{mnpq}=\bar{\t{\theta}}{\g}_{mnpq}\t{\theta} \label{eq:4form}.
\end{equation}
From (\ref{eq:cond2b}) we see
\begin{equation}
\phi_{pqr}G^{mpqr}=0 \label{eq:compG}.
\end{equation}
This equation gives us the non-zero components of $G$ depending on the
choice of the seven-manifold. Equation (\ref{eq:cond1b}) then encodes
the relationship between the dual 4-form, $*\phi$, and the non-zero
components of $G$.
We now do a case by case study of the
various possibilities for ${X}^7$ classified by their holonomy
groups ${\mathcal H}$. In the following subsections we work
exclusively with tangent space indices which we denote by a prime
($\;'\;$) on the index. For example, an orthonormal basis
will be denote by $\{e_{a'}\}$. The constructions that follow are very
similar to what was done without a warp-factor in \cite{Pope}.
\subsection{${\mathcal H}=SU(3)$}
We set $X^6=CY_3$. On a $CY_3$
$\phi$ is given by 
\begin{equation}
\phi=\omega\w e_{7'}+{\rm Im}(\Omega)
\end{equation}
where $\omega$ and $\Omega$ are the K\"ahler 2-form and the
$SU(3)$-structure 3-form, respectively. They are given by
\begin{eqnarray*}
\omega=e_{1'}\w e_{2'}+e_{3'}\w e_{4'}+e_{5'}\w e_{6'} \nonumber
\end{eqnarray*}
\begin{eqnarray*}
\Omega=\left(e_{1'}+ie_{2'}\right)\w\left(e_{3'}+ie_{4'}\right)\w\left(e_{5'}+ie_{6'}\right).
\end{eqnarray*} We can see from (\ref{eq:compG}) that only
seven components of $G$ can be non-zero.
\subsection{${\mathcal H}=SU(2)$}
We can take $X^6=K3\times T^2$. Then $\phi$ is
given by
\begin{eqnarray}
&\phi=\omega_{1'}\w e_{5'}+\omega_{2'}\w e_{6'}+\omega_{3'}\w
e_{7'}+e_{5'}\w e_{6'}\w e_{7'}\nonumber \\
&+ {\rm \;cyclic\; perm.\; of\; e_{5'},e_{6'}\;and\;e_{7'}\;in\;the\;first
\;three\;terms}.
\end{eqnarray}
where $\omega_{1'}$ is the K\"ahler 2-from and $\omega_{2'}+i\omega_{3'}$
is the complex volume 2-form of $K3$. They are given by
\begin{eqnarray*}
\omega_{1'}=e_{1'}\w e_{4'}+e_{2'}\w e_{3'} \\
\omega_{2'}=e_{1'}\w e_{3'}-e_{2'}\w e_{4'} \\
\omega_{3'}=e_{1'}\w e_{2'}+e_{3'}\w e_{4'}.
\end{eqnarray*}
We can now see from (\ref{eq:compG}) that only one component of $G$ can be non-zero and it
the one proportional to the volume form of $K3$.


\subsection*{\it Shape of the Warp-Factor}
\begin{figure}[t]
\epsfysize=8cm
\centerline{\epsffile{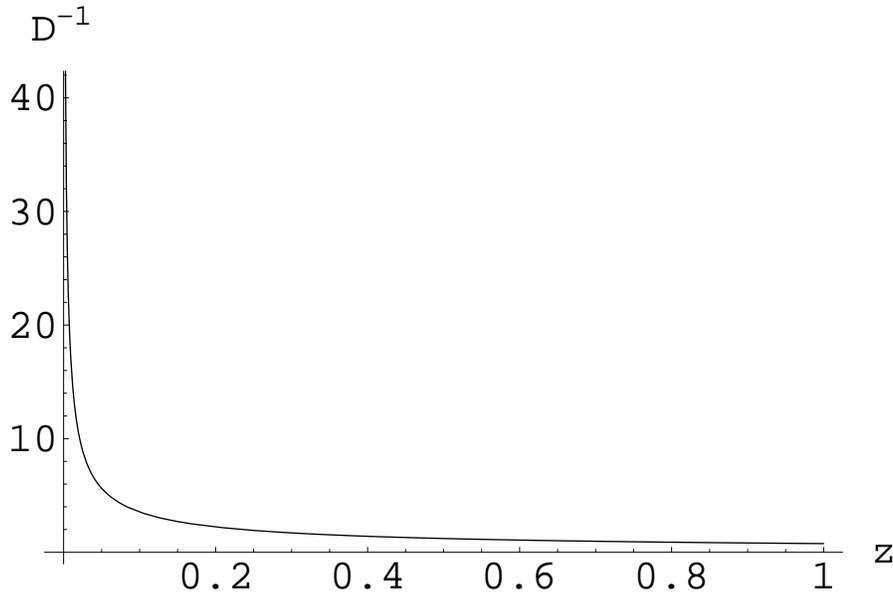}}
\caption{Shape of the warp-factor.} 
\end{figure}
If we have assumed that there are no M2- or M5-brane sources then
supersymmetry restricts us to look for solutions of the form
\begin{equation}
d\b{s}^2=D^{-1}(z)\;ds^2_{\bbbR^{1,3}}+D(z)\;\{d
z^2+ds^2_{X^6}\}.
\end{equation}
Einstein's equation then reduces to
\begin{equation}
\fc{3}{2}\;\left(\fc{d\log{D}}{dz}\right)^{2}=-\fc{d^2\log{D}}{dz^2}
\label{eq:shapediff}
\end{equation}
Eq.(\ref{eq:shapediff}) is solved easily to give
\begin{equation}
D(z)=e^{a_{2}}\left\vert a_{1}-\fc{3}{2}\; z
\right\vert^{\fc{2}{3}} \label{eq:shape}
\end{equation} 
where $a_1$ and $a_2$ are integration constants. The metric is
singular at $z=\fc{2a_{1}}{3}$.
The singularity, however, is repulsive for massive particles.

\section{Discussion}
So we have seen that to have warped compactification of M-theory down
to four-dimensional Minkowski space with $G$-fluxes on the internal
space we need to consider more complicated spinor \an than usually
assumed. The fact that our solution is singular is not surprising in
the context of the no-go theorem of \cite{Gibbons2,Maldacena}. However, the
reducibility of the internal space in our case is due to the fact that
our spinor \an contains a covariantly constant vector field. An \an
with a covariantly constant 3-form field should then be the
appropriate \an for $G_2$-holonomy compactification. For such
compactifications one should allow for singular internal spaces (and
thus possibly side step the no-go theorem) of the
type explored in \cite{Acharya2} to generate chiral fermions.
\par
A possible application of the simple model discussed here could be in
the context of Ho\v rava-Witten theory. There  \11 lives
in the bulk and one of the direction is a line which terminates on
10-{\it d} ``branes''. An observer living in the bulk sees usual \11
but the physics on the branes are quite different, containing chiral
multiplets transforming under $E_8$ or $\bar{E_8}$. So it should be
possible to embed our solution in the bulk of Ho\v rava-Witten
theory and put one the branes before one can reach the
singularity. In our solution 6 of the dimensions have been
compactified and so in effect we have a theory where there are two
4-{\it d} Minkowski branes living on the ends of a line. Since the
metric is warped it is clear that observers on each of these branes
will observe different physics related to the ``fifth''
dimension. This is a supersymmetric analog of the much studied
brane-world scenarios \cite{Randall}.

\begin{center}
\section*{\small ACKNOWLEDGEMENTS}
\end{center}
I would like to thank C. Codirla, G.W. Gibbons, M.J. Perry,
D.J. Scruton and J.F. Sparks for useful discussions. This work was partly funded by a Cambridge Philosophical Society grant.
\appendix
\section{Dirac Matrix Identities} \label{ap:a}
\setcounter{equation}{0}
We adopt the convention in which 11-{\it d} space-time has signature
$\{-++...+\}$. The time-like Dirac matrix is
anti-hermitian and the rest are hermitian. They obey
\begin{equation}
\{\Gamma_{M},\Gamma_{N}\}=2g_{MN}\bbbone.
\end{equation}
The Ricci identity for spinors is
\begin{equation}
\left[\na{M},\na{N}\right]\psi=\fc{1}{4}R_{MNPQ}\Gamma^{PQ}\psi
\end{equation}
and spheres in our convention have positive curvature.
Decomposition of the Dirac matrices appropriate for four-seven split
is
\begin{eqnarray}\begin{array}{rcl}
\Gamma_{\mu} & = & \gamma_{\mu}\otimes \bbbone \\
\Gamma_{m} & = & \Sigma^{5}\otimes \gamma_{m} \end{array}
\end{eqnarray}
where we have defined
\begin{equation}
\Sigma^{5}=\fc{i}{4!}\eps_{\mu\nu\rho\sigma}\g^{\mu}\g^{\nu}\g^{\rho}\g^{\sigma}.
\end{equation}
$\eps_{\mu\nu\rho\sigma}$ is the 4-{\it d} Levi-Civita tensor
density. In the text we also make use of
\begin{equation}
\g^{\nu\rho\sigma}=i\Sigma^{5}\g_{\mu}\eps^{\nu\rho\sigma\mu}
\end{equation}
and so
\begin{equation}
\eps_{\mu\nu\rho\sigma}\g^{\nu\rho\sigma}=6i\Sigma^{5}\g_{\mu}
\end{equation}
\par
Let us now turn to an important issue about Majorana spinors. On
7-{\it d} Euclidean manifolds, it only
possible to define {\it pseudo}-Majorana spinors which means that the charge conjugation
matrix is symmetric \cite{Salam}. The transpose of the Dirac matrices are given by
\begin{equation}
\g^{T}_{m}=-C\g_{m}C^{-1}
\end{equation}
and the pseudo-Majorana condition is then given by
\begin{equation}
\b{\theta}=\theta^{T}C.
\end{equation}
If $\theta$ is a {\it commuting} pseudo-Majorana spinor then we easily
obtain 
\begin{equation}\begin{array}{rcl}
\b{\theta}\g^{m_{1}...m_{n}}\theta&=&\theta^{T}C\g^{m_{1}...m_{n}}\theta
\\
&=&\left(-1\right)^{\fc{n^{2}+n}{2}}\b{\theta}\g^{m_{1}...m_{n}}\theta
\end{array}
\end{equation}
so we see that the antisymmetric tensor
$\b{\theta}\g^{m_{1}...m_{n}}\theta$ vanishes except for
$n=0,3,4,7$. In a similar manner it  can be shown that the analogous
tensor in 11-{\it d} Lorentzian signature space made out of an {\it
anticommuting} Majorana spinor $\eta$ vanishes except for $n=0,3,4,7,8,11$.
\par
A useful identity that lets us expand the product of two elements of
the Clifford algebra as a linear combination is
\begin{eqnarray}
\Gamma_{M_{1}...M_{m}}\Gamma^{N_{1}...N_{n}}={\Gamma_{M_{1}...M_{m}}}^{N_{1}...N_{n}}+mn{\Gamma_{M_{1}...M_{m-1}}}^{N_{2}...N_{n}}{\delta_{M_{m}}}^{N_{1}}\\
\nonumber
+\fc{1}{2!}m(m-1)n(n-1){\Gamma_{M_{1}...M_{m-2}}}^{N_{3}...N_{n}}{\delta_{M_{m-1}}}^{N_{1}}{\delta_{M_{m}}}^{N_{2}}+...
\end{eqnarray}
from which one can derive the following identities for the
seven-dimensional Dirac matrices,
\begin{equation}\begin{array}{cccccc}
\left[\g_{m},\g^{n}\right]&=&2{\g_{m}}^{n}&\{\g_{m},\g^{n}\}&=&2{\d_{m}}^{n}
\\
\\
\{{\g_{mn}},\g^{r}\}&=&2{\g_{mn}}^{r}&\left[\g_{mn},\g^{r}\right]&=&-4{\d^{r}}_{[m}\g_{n]}\\
\\
\left[\g_{mnp},\g^{r}\right]&=&2{\g_{mnp}}^{r}&\{\g_{mnp},\g^{r}\}&=&6{\d^{r}}_{[m}\g_{np]}\\
\\
\{\g_{mnpq},\g^{r}\}&=&2{{\g_{mnpq}}^{r}}&\left[\g_{mnpq},\g^{r}\right]&=&-8{\d^{r}}_{[m}\g_{npq]}\\
\\
\left[\g_{mnpqk},\g^{r}\right]&=&2{\g_{mnpqk}}^{r}&\{\g_{mnpqk},\g^{r}\}&=&10{\d^{r}}_{[m}\g_{npqk]}.
\end{array}
\end{equation}
In our convention $\bbbone, \g^m, \g^{mnpq}, \g^{mnpqr}$ are hermitian while the
rest of the elements of the Clifford algebra on the internal space are
antihermitian.


\end{document}